\documentclass[10pt]{article}
\usepackage{amssymb}
\usepackage{amsfonts}
\usepackage{amsmath}
\usepackage{graphicx}
\usepackage{indentfirst} 
\usepackage{cite} 

\long\def\ca#1\cb{} 

\newcommand{\ad}{^\dagger }

\newcommand{\becs}{\begin{cases}}
\newcommand{\bem}{\begin{matrix}}

\newcommand{\bra}[1]{\langle#1|}

\newcommand{\dya}[1]{|#1\rangle\langle#1|}
\newcommand{\dyad}[2]{|#1\rangle\langle#2|}

\newcommand{\encs}{\end{cases}}
\newcommand{\enm}{\end{matrix}}

\newcommand{\hquad}{\mspace{8 mu}} 
\newcommand{\inp}[1]{\langle#1|#1\rangle }
\newcommand{\inpd}[2]{\langle#1|#2\rangle }
\newcommand{\ket}[1]{|#1\rangle }

\newcommand{\mat}[1]{\left(\begin{matrix}#1\end{matrix}\right)}

\newcommand{\mted}[3]{\langle#1|#2|#3\rangle }

\newcommand{\ot}{\otimes }

\newcommand{\pr}{\partial }
\newcommand{\ra}{\rightarrow }

\renewcommand{\Re}{\hbox{Re}} 

\newcommand{\tm}{\times }
\newcommand{\Tr}{{\rm Tr}}
\newcommand{\trp}{^\textup{T}}

\newcommand{\BC}{{\mathcal B}}
\newcommand{\CC}{{\mathcal C}}
\newcommand{\DC}{{\mathcal D}}

\newcommand{\HC}{{\mathcal H}}

\newcommand{\JC}{{\mathcal J}}
\newcommand{\KC}{{\mathcal K}}

\newcommand{\QC}{{\mathcal Q}}

\newcommand{\TC}{{\mathcal T}}

\newcommand{\al}{\alpha }
\newcommand{\bt}{\beta }
\newcommand{\gm}{\gamma }

\newcommand{\dl}{\delta }
\newcommand{\Dl}{\Delta }

\newcommand{\zt}{\zeta }

\newcommand{\lm}{\lambda }

 \def\outl#1{}  \def\xa{} \def\xb{}  

\begin{document}
\title{Degradable Quantum Channels Using Pure to Product of Pure States
  Isometries}
\author{Vikesh Siddhu\thanks{vsiddhu@andrew.cmu.edu} \
 and Robert B. Griffiths\thanks{rgrif@cmu.edu}\\
Department of Physics,
Carnegie-Mellon University,\\
Pittsburgh, PA 15213, USA}
\date{Version of 23 November 2016}
\maketitle

\begin{abstract}
  We consider a family of quantum channels characterized by the fact that
  certain (in general nonorthogonal) Pure states at the channel entrance are
  mapped to (tensor) Products of Pure states (PPP, hence ``pcubed'') at the
  complementary outputs (the main output and the ``environment'') of the
  channel. The pcubed construction, a reformulation of the twisted-diagonal
  procedure by M.~M.~Wolf and D.~ {P\'erez-Garc\'\i a} [Phys.\ Rev.\ A 75,
  012303 (2007)], can be used to produce a large class of degradable quantum
  channels; degradable channels are of interest because their quantum
  capacities are easy to calculate. Several known types of degradable channels
  are either pcubed channels, or subchannels (employing a subspace of the
  channel entrance), or continuous limits of pcubed channels. The pcubed
  construction also yields channels which are neither degradable nor
  antidegradable (i.e., the complement of a degradable channel); a particular
  example of a qutrit channel of this type is studied in some detail.
  Determining whether a pcubed channel is degradable or antidegradable or
  neither is quite straightforward given the pure input and output states that
  characterize the channel. Conjugate degradable pcubed channels are always
  degradable.
\end{abstract}

\tableofcontents

\section{Introduction}
\label{sct1} 

The transmission of quantum information is a central problem in quantum
information theory, and a great deal of attention has been devoted to
understanding noisy quantum channels, of which the simplest kind is defined in
Sec.~\ref{sbct2.1}. Of particular interest is finding the asymptotic quantum
capacity $Q$ of such a channel, a task which is much more difficult than its
analog in classical information theory. The asymptotic capacity of a classical
channel is equal to the mutual information between the channel's input and
output (see Ch.~7 in \cite{CvTh06}), maximized over all possible probability
distributions for the input. Because the mutual information is a concave
function of the input probability distribution, this quantity is easily
computed numerically if a closed-form expression is not available. In
addition the classical capacity is additive: the capacity of two independent
channels used in parallel is just the sum of the individual capacities.

A quantum analog of the classical mutual information is the coherent
information, and finding its maximum over all possible input density operators
yields the \emph{one-shot quantum capacity} $Q^{(1)}$ (sometimes also called
the coherent information) of the channel (see Ch.~13 in \cite{Wld16}).
But unlike its classical counterpart, the coherent information need
not be a concave function of the input density operator, and $Q^{(1)}$ is in
general not additive for two channels placed in parallel. (See
Sec.~\ref{sbct2.2} for additional remarks.)

However, Devetak and Shor \cite{DvSh05} showed that $Q^{(1)}$ is additive for
any two (not necessarily identical) \emph{degradable} quantum channels (see the
definition in Sec.~\ref{sbct2.3}) placed in parallel. From this it follows that
$Q=Q^{(1)}$ for a degradable channel, and $Q$ for two such channels placed in
parallel is the sum of its values for the individual channels. In addition, the
coherent information of a degradable channel is a concave function of the input
density operator~\cite{YrHD08}, allowing $Q^{(1)}$ to be computed
efficiently. In this respect the quantum capacity problem for degradable
quantum channels resembles that of classical channels. However, checking
whether a given quantum channel is or is not degradable is not altogether
straightforward, and while several interesting classes of degradable channels
are known \cite{WlPG07,CbRS08} the full limits of this family of channels have
not been established.

Very little is known about the asymptotic capacities of quantum channels which
are not degradable. Cases are known for which $Q^{(1)}$ and $Q$ are not
additive; one of the most striking \cite{SmYr08} is a pair of channels both of
which have asymptotic capacities $Q$ equal to zero, but when placed in parallel
have a finite $Q^{(1)}$, and thus a nonzero asymptotic $Q$. Various upper and
lower bounds on capacities of nondegradable channels are known---some obtained
using a clever construction involving degradable channels \cite{SmSm08}---but
at present the source of nonadditivity is not understood. One may hope that a
better understanding of degradable channels, including what happens when
degradability breaks down upon variation of some parameter, may provide useful
insights.

In this article we show that a large class of degradable channels can be
constructed using isometries which map Pure states to Products of Pure states,
thus PPP or \emph{pcubed}. Not all pcubed channels are degradable, but testing
for degradability is relatively easy, and degradable families constructed in
this way include several examples known previously. Indeed, by including
subchannels (which use a subspace of the channel entrance), and continuous
limits, as illustrated in some examples in Sec.~\ref{sct4}, it seems
possible to provide pcubed constructions for many of the degradable channels
discussed in \cite{CbRS08}, although we have not studied every case found in
that very useful paper. Not all pcubed channels are degradable, and it is easy 
to construct examples in which a continuous variation of a parameter leads
from a degradable to a nondegradable channel, thus allowing a study of what
happens when the pleasant properties associated with degradability are lost.
Section~\ref{sct5} is devoted to a very simple example. 

The pcubed construction is in effect what Chefles and his collaborators used
when considering the problem of quantum channels that map certain sets of pure
states to other sets of pure states \cite{Chfl00,Chfl02,ChJW04}. And it
coincides in many cases, see Sec.~\ref{sbct3.2}, with the \emph{twisted
  diagonal} construction used by Wolf and P\'erez-Garc\'ia \cite{WlPG07}
in their study of degradable channels. The main way our work differs from 
previous studies is in focusing on the isometry which generates a channel
superoperator, rather than on the superoperator itself. The two are formally
equivalent, but sometimes the isometry is easier to understand; in
particular when studying degradable channels, but perhaps in other cases as
well.

The remainder of this article is structured as follows. Section~\ref{sct2}
contains various definitions and relationships which, while not new, should
help make the present treatment reasonably self contained. In particular, the
relationship between isometries, Kraus operators, and channel superoperators is
summarized in Sec.~\ref{sbct2.1}, coherent information and the quantum
capacities $Q^{(1)}$ and $Q$ are defined in Sec.~\ref{sbct2.2}, and degradable
and conjugate degradable \cite{BDHM10} channels in Secs.~\ref{sbct2.3} 
and~\ref{sbct2.4}. Section~\ref{sct3} introduces pcubed isometries and channels.
Following the basic definitions in Sec.~\ref{sbct3.1}, the connection with
twisted-diagonal channels is worked out in Sec.~\ref{sbct3.2}, and degradable
pcubed channels are discussed in Secs.~\ref{sbct3.3}. An argument in
Sec.~\ref{sbct3.4} shows that a pcubed channel which is conjugate degradable is
also degradable. Specific examples of degradable pcubed channels are discussed
in Sec.~\ref{sct4}; these include some qubit and qutrit channels, erasure
channels, and Hadamard channels (the complements of entanglement-breaking
channels). Section~\ref{sct5}, with some details in the Appendix, presents
results for a particular qutrit channel in which a continuous variation of
parameters leads out of a degradable regime, and certain properties associated
with a degradable channel break down.  Section~\ref{sct6}
summarizes the paper.

\section{Quantum Isometries and Channels}
\label{sct2}

\subsection{Channels and complementary channels}
\label{sbct2.1}

\xb \outl{Qm channel, channel entrance, isometry, complementary channels,
  superoperators, Kraus operators} \xa

A noisy quantum channel, also called a quantum operation, can be constructed
starting with a unitary operator $U$, representing the time development of a
closed quantum system, that carries $\HC_a\ot\HC_e$ to $\HC_b\ot\HC_c$, where
the Hilbert space $\HC_a$ represents the \emph{channel entrance}, $\HC_e$ the
environment, and $\HC_b$ and $\HC_c$ the \emph{direct} and \emph{complementary}
channel outputs. One often equates $\HC_b$ with $\HC_a$ and $\HC_c$ with
$\HC_e$, i.e., the complementary output is the environment, but greater
generality is possible if one allows the dimensions of $\HC_a$ and $\HC_b$ to
be different. If the environment is always in the same initial state
$\ket{e_0}$ (a mixed state of the environment can always be purified), its
action on the states in $\HC_a$ can be represented by an \emph{isometry}
\begin{equation}
  J:\HC_a \ra \HC_b \ot \HC_c;\quad J\ket{a} := U\ket{a}\ot\ket{e_0}.
\label{eqn1}
\end{equation}
Our discussion is based on this isometry, and assumes the three Hilbert spaces
have (finite) dimensions $d_a$, $d_b$, and $d_c$, which can be arbitrary except
for the requirement $d_a \leq d_b d_c$, which is necessary for $J$ to be an
isometry, i.e., $J\ad J=I_a$. The isometry then produces two quantum
channels represented by the superoperators
\begin{equation}
 \BC(O) = \Tr_c(J O J\ad),\quad \CC(O) = \Tr_b(J O J\ad)
\label{eqn2}
\end{equation}
carrying the space $\hat\HC_a$ of operators on $\HC_a$, of which $O$ is a
typical example, into the operator spaces $\hat\HC_b$ and $\hat\HC_c$,
respectively. It is customary to refer to $\CC$ as the \emph{complement} of
$\BC$, with $\BC$ the complement of $\CC$, so a single isometry produces a
complementary pair of channels. The fact that $J$ is an isometry implies that
$\BC$ and $\CC$ are \emph{completely positive and trace preserving} (CPTP);
conversely, any CPTP map can be produced in this way starting with a suitable
isometry. One way to represent the map $\BC$ is to use Kraus operators $L_l:
\HC_a \ra \HC_b$ which define the isometry
\begin{equation}
 J = \sum_l \ket{c_l} \ot L_l,\quad \BC(O) = \sum_l L_l^{} O L_l\ad,
\label{eqn3}
\end{equation}
where $\{\ket{c_j}\}$ is an orthonormal basis of $\HC_c$. Of course a similar
representation is possible for $\CC$, with Kraus operators mapping $\HC_a$ to
$\HC_c$.

\xb
\outl{Define \emph{image subspace} and \emph{subchannel} }
\xa

The isometry $J$ maps the channel entrance $\HC_a$ onto a subspace of
$\HC_b\ot\HC_c$, and this \emph{image subspace} determines the properties of
the channel pair up to a unitary on $\HC_a$; i.e., if an isometry $J'$ has the
same image subspace as $J$, then $J' = JU$ for some unitary $U$ on $\HC_a$. If
$J$ itself is restricted to a subspace of $\HC_a$, then it is (obviously) again
an isometry acting on this restriction, and produces a pair of channels which
we will refer to as \emph{subchannels} of the original channels; equivalently,
these subchannels are associated with a subspace of the original image subspace
on $\HC_b\ot\HC_c$.

\xb
\outl{Channels in parallel}
\xa

Two quantum channels $\BC_1$ and $\BC_2$ are said to be ``in parallel'' or
``used simultaneously'' or ``used independently'' if together they constitute a
single channel $\BC$ defined as follows. If $J_1:\HC_{a1}\ra
\HC_{b1}\ot \HC_{c1}$ and $J_2:\HC_{a2}\ra \HC_{b2}\ot \HC_{c2}$ are the
isometries that lead to $\BC_1$ and $\BC_2$, let $J = J_1\ot J_2$ be the
map from $(\HC_a=\HC_{a1}\ot\HC_{a2})$ to $(\HC_b= \HC_{b1}\ot\HC_{b2})\ot
(\HC_c=\HC_{c1}\ot\HC_{c2})$ 
\ca 
$\ot\HC_c$, with
$\HC_b=\HC_{b1}\ot\HC_{b2}$ and $\HC_c=\HC_{c1}\ot\HC_{c2}$, defined in the
obvious way: 

$J:\HC_a \ra \HC_b\ot\HC_c$, where $\HC_a=\HC_{a1}\ot\HC_{a2}$
and $\HC_b=\HC_{b1}\ot\HC_{b2}$,
\cb
defined in the obvious way:
\begin{equation}
 J(\ket{a1}\ot\ket{a2}) = (J_1\ket{a1})\ot (J_2\ket{a2}).
\label{eqn4}
\end{equation}
Then $\BC=\BC_1\ot\BC_2$ as defined in \eqref{eqn2} for $J$ in \eqref{eqn4} is
what we mean by $\BC_1$ and $\BC_2$ in parallel, and of course its complement
$\CC$ is $\CC_1$ and $\CC_2$ in parallel. The definition extends in an obvious
way to three or more channels in parallel.

\subsection{Coherent information and quantum capacity}
\label{sbct2.2}

The \emph{coherent information} or \emph{entropy bias} of a channel $\BC$ with
complementary channel $\CC$ is defined by
the formula
\begin{equation}
 \Dl S(\rho) = S(\BC(\rho)) - S(\CC(\rho)),
\label{eqn5}
\end{equation}
where $\rho$ is the input density operator, a positive operator with trace
equal to 1, and $S(\rho)= -\Tr(\rho\log\rho)$ is the von Neumann entropy. The
\emph{one-shot quantum capacity}, also known (somewhat confusingly) as the
\emph{coherent information of the channel}  is defined by the formula
\begin{equation}
 Q^{(1)}(\BC) = \max_{\rho} \Dl S(\rho),
\label{eqn6}
\end{equation}
where the maximum is taken over all density operators in $\hat\HC_a$.
The one-shot capacity for the $\CC$ channel can be written as 
\begin{equation}
 Q^{(1)}(\CC) = - \min_{\rho} \Dl S(\rho).
\label{eqn7}
\end{equation}
The one shot capacity is superadditive, meaning that for channels $\BC_1$ and
$\BC_2$ in parallel,
\begin{equation}
 Q^{(1)}(\BC_1\ot\BC_2) \geq Q^{(1)}(\BC_1)+ Q^{(1)}(\BC_2). 
\label{eqn8}
\end{equation}
The (asymptotic) quantum capacity $Q(\BC)$ \cite{KrWr04,Smth10}
is equal to
\begin{equation}
 Q(\BC) = \lim_{n \mapsto \infty} \frac{1}{n} Q^{(1)}(\BC^{\otimes n}),
\label{eqn9}
\end{equation}
and thus in light of \eqref{eqn8} is bounded below by $Q^{(1)}(\BC)$.

\subsection{Degradable channels}
\label{sbct2.3}

\xb
\outl{Define degradable (pair), antidegradable, using isometry $K$}
\xa

One of the channels in a pair produced by an isometry is said to be
\emph{degradable} \cite{DvSh05} if its output can be sent through another noisy
channel in such a way that the combination is equivalent to the other,
complementary, channel in the pair, which is then said to be
\emph{antidegradable}. We shall also refer to a \emph{degradable pair} of
channels, and in addition use the term \emph{undegradable} to refer to either a
single channel or a complementary pair in a situation in which the channel
is neither degradable nor antidegradable, and thus the same is true of the
complementary channel.

In particular, the $a\ra b$ channel of Sec.~\ref{sct2}, with
superoperator $\BC$ defined in \eqref{eqn2} is degradable if there is an
additional isometry
\begin{equation}
K:\HC_b \ra \HC_c \ot \HC_d
\label{eqn10}
\end{equation}
giving rise to a superoperator
\begin{equation}
 \DC(S) = \Tr_d(K S K\ad)
\label{eqn11}
\end{equation}
 with the property that, for $\CC$ defined in \eqref{eqn2},
\begin{equation}
 \CC = \DC \circ \BC,\text{ i.e., } \CC(O) = \DC(\BC(O)).
\label{eqn12}
\end{equation}
The ancillary space $\HC_d$ plays no further role in the definition of
degradability. One can also think of $K$ as mapping to a new
Hilbert space $\HC'_c$ isomorphic to $\HC_c$, but the simplest statement of the
degradability condition uses $\HC_c$. Note that it is possible for a channel to
be simultaneously degradable and antidegradable, in which case its complement
is also both degradable and antidegradable.

\xb
\outl{Additivity of coherent info, Qm capacity for degradable channels}
\xa

If two degradable channels $\BC_1$ and $\BC_2$ are placed in parallel, the
combined channel $\BC_1\ot\BC_2$ is again degradable, and in addition the
inequality \eqref{eqn8} becomes an equality \cite{DvSh05}. This has the
consequence that for any degradable channel $\BC$ the asymptotic capacity
$Q(\BC)$ is equal to $Q^{(1)}(\BC)$, and if both $\BC_1$ and $\BC_2$ are
degradable, $Q(\BC_1\ot\BC_2)$ is equal to $Q(\BC_1)+Q(\BC_2)$. In addition
$\Dl S(\rho)$, \eqref{eqn5}, is a concave function of $\rho$ when $\BC$ is
degradable \cite{YrHD08}, so its maximum $Q^{(1)}$ is relatively easy to
calculate. As a consequence, quantum capacities of degradable channels inherit
some of the nice properties of Shannon capacities of classical noisy channels.

\subsection{Conjugate degradable channels}
\label{sbct2.4}

A quantum channel $\BC$ with complement $\CC$ is said to be \emph{conjugate
  degradable} \cite{BDHM10} provided there is a CPTP map $\bar\DC$ from
$\hat\HC_b$ to $\hat\HC_c$ such that
\begin{equation}
 \TC\circ\CC = \bar\DC \circ \BC,
\label{eqn13}
\end{equation}
where $\TC$ is a superoperator defined by choosing a particular orthonormal
basis $\{\ket{c_k}\}$ of $\HC_c$, and for any operator $M\in \hat\HC_c$
defining $M\trp=\TC(M)$ as the operator whose matrix is related to that of $M$
by a transpose:
\begin{equation}
\mted{c_l}{M\trp}{c_k} = \mted{c_k}{M}{c_l}.
\label{eqn14}
\end{equation}
A channel can be both degradable and conjugate degradable if both \eqref{eqn12}
and \eqref{eqn13} are satisfied, which will in general require distinct
maps $\DC$ and $\bar\DC$. 

One can also define conjugate degradability by replacing $\TC$ in \eqref{eqn13}
with a complex conjugation operator $\KC$, defining $M^\textup{C} = \KC(M)$ to
be the operator such that
\begin{equation}
 \mted{c_l}{M^\textup{C}}{c_k} = \mted{c_l}{M}{c_k}^*.
\label{eqn15}
\end{equation}
Note that $\KC$ is an antilinear operator, and for this reason \eqref{eqn13}
with $\KC$ in place of $\TC$ is only required to hold for Hermitian operators
$O=O\ad$, which of course includes density operators. Because $\TC$ is linear,
if \eqref{eqn13} holds for Hermitian operators it also holds for all operators
in $\hat\HC_a$.

Both $\TC$ and $\KC$ are defined for a particular orthonormal basis of $\HC_c$,
but the basis dependence is of no importance for the present discussion; if a
different basis is used, one only needs to adjust $\CC$ and $\bar\DC$ by
applying a suitable unitary (i.e., $\rho\ra U\rho U\ad$) to their outputs, and
\eqref{eqn13} will again be satisfied.

At present it is not known if there are channels which are conjugate degradable
but not degradable. If they exist they are \emph{not} of the pcubed type, see
Sec.~\ref{sbct3.4}.

\section{Pcubed Isometries and Channels}
\label{sct3}

\subsection{Definitions}
\label{sbct3.1}

\xb
\outl{Kets that define the isometry. $A$, $B$, $C$ matrices; $A=B*C$}
\xa

A pure state to product of pure states, or pcubed, isometry $J$ is defined as 
follows.  Let $\{\ket{\al_j}\}$ for $1\leq j \leq d$ be collection of
normalized kets that span $\HC_a$, and $\{\ket{\bt_j}\}$ and $\{\ket{\gm_j}\}$
be collections of normalized kets in $\HC_b$ and $\HC_c$ such that
\begin{equation}
 J \ket{\al_j} = \ket{\bt_j}\ot\ket{\gm_j} \text{ for } 1\leq j \leq d.
\label{eqn16}
\end{equation}
The necessary and sufficient condition for $J$ to be an isometry
is that it preserves inner products, which is to say:
\begin{equation}
 A_{jk} = \inpd{\al_j}{\al_k} = \inpd{\bt_j}{\bt_k}\inpd{\gm_j}{\gm_k}
= B_{jk} C_{jk},\text{ or } A = B*C,
\label{eqn17}
\end{equation}
where we use $*$ to denote the element-wise or Hadamard product of the Gram
matrices $A$, $B$, and $C$ constructed from the three collections of kets. Note
that the diagonal elements of these matrices are all equal to 1 by the
normalization condition, and since they are Hermitian, \eqref{eqn17} constitutes
a set of $d(d-1)/2$ independent equations for the (in general complex)
off-diagonal elements.

\xb \outl{Kets $\{\ket{\al_j}\}$, etc. need not be orthogonal. Relationship
  between $J$ and $A,B,C$ not unique} \xa

The kets in $\{\ket{\al_j}\}$ are \emph{not} assumed to be orthogonal, and they
need not be linearly independent as long as their span is equal to $\HC_a$, so
that the isometry in \eqref{eqn16} is well defined. The same is true of the
collection $\{\ket{\bt_j}\}$, except that it need not span $\HC_b$, and
similarly the $\{\ket{\gm_j}\}$ need not span $\HC_c$. In applications we
generally assume that the $\{\ket{\al_j}\}$ are linearly independent, and thus
form a (in general not orthonormal) basis of $\HC_a$, so $d=d_a$; and also that
$d_b=d_c=d$, though the $\{\ket{\bt_j}\}$ and $\{\ket{\gm_j}\}$ need not span
$\HC_b$ and $\HC_c$.
The isometry $J$ in \eqref{eqn16} is not changed if for each $j$, $\ket{\al_j}$,
$\ket{\bt_j}$ and $\ket{\gm_j}$ are multiplied by phases $e^{i\lm_j}$,
$e^{i\mu_j}$ and $e^{i\nu_j}$, with $\lm_j-\mu_j-\nu_j$ a multiple of $2\pi$.
However, this will in general change the off-diagonal elements of $A$, $B$,
and $C$, though \eqref{eqn17} will still be satisfied. Conversely, a set of $A$,
$B$, and $C$ matrices satisfying \eqref{eqn17} determine $J$ only up to local
unitaries, since a Gram matrix only determines the corresponding kets up to a
unitary transformation.

\xb
\outl{Properties of Gram matrices. Using $B$, $C$ or $A$, $B$ to construct $J$}
\xa

As $A$, $B$ and $C$ are Gram matrices, they are positive semi-definite, and if
the $\ket{\al_j}$ are linearly independent, $A$ will be positive definite, with
all its eigenvalues strictly positive. The Hadamard product of two positive
definite matrices is positive definite (p.~458 of \cite{HrJh85}),
whereas if only one of the matrices is positive definite and the other is
positive semidefinite, or both are positive semidefinite, the product will
certainly be positive semidefinite, but may or may not be positive definite.
Since any positive semidefinite matrix is a Gram matrix for a suitable (not
unique) collection of kets, what we call a pcubed isometry can be obtained by
choosing  any two positive (semi)definite matrices $B$ and $C$, with $1$'s
on the diagonal, and taking the product \eqref{eqn17}. Or one can start with a
positive (semi)definite $A$ and a positive (semi)definite $B$ and define
$C=A/*B$ by element-wise division; if $C$ is positive semidefinite, the three
matrices define a pcubed isometry. However, if for some $j$ and $k$ both
$A_{jk}$ and $B_{jk}$ are zero, the quotient is not well defined, and there may
or may not be values of $C_{jk}$ for which $C$ has the desired positivity
property. (In our discussion of degradable channels in Sec.~\ref{sbct3.3} we
will assume that both the $\ket{\al_j}$ and the $\ket{\bt_j}$ are linearly
independent bases of $\HC_a$ and $\HC_b$.)

\xb
\outl{The $J\HC_a$ subspace of $\HC_b\ot\HC_c$ is spanned by products} 
\xa

The image subspace (see the definition in Sec.~\ref{sct2}) of a pcubed isometry
is spanned by products of pure states $\ket{\bt_j}\ot\ket{\gm_j}$, and such
subspaces are rather special. On the other hand, since the entire space
$\HC_b\ot\HC_c$ is spanned by such products, any subspace is always a subspace
of one of these ``special'' types of spaces---which means that only in
exceptional circumstances will subchannels of pcubed channels be of interest.
One of these exceptional circumstances is that in which the $a\ra b$ channel
$\BC$ is degradable, in which case the corresponding subchannel is also
degradable.

\subsection{Twisted diagonal channels}
\label{sbct3.2}

A twisted-diagonal $a\ra b$ channel is defined in \cite{WlPG07} for the case
$d_a=d_b=d$ through the requirement that the Kraus operators
$L_l:\HC_a\ra \HC_b,\,\,1\leq l\leq d,$ be simultaneously diagonalizable in the
sense that given orthonormal bases $\{\ket{a_j}\}$ of $\HC_a$ and
$\{\ket{b_k}\}$ of $\HC_b$, there are invertible operators $X\in\hat\HC_a$ and
$Y\in\hat\HC_b$ such that these make all the Kraus operators diagonal:
\begin{equation}
 \mted{b_k}{Y L_l X}{a_j} = \lm_{jl} \dl_{jk}. 
\label{eqn18}
\end{equation}
This means there is a collection of $d$ linearly independent ($X$ is
nonsingular) kets $\ket{\bar a_j}=X\ket{a_j}$ in $\HC_a$ and a corresponding
linearly independent collection of kets $\ket{\bar b_j}$ in $\HC_b$ such that
every Kraus operator maps every $\ket{\bar a_j}$ to a (possibly zero) multiple
of $\ket{\bar b_j}$. The formal argument showing that the corresponding
isometry is pcubed begins by noting that \eqref{eqn18} implies that
\begin{equation}
 L_l = \sum_j \lm_{jl} \bar Y \dyad{b_j}{a_j} \bar X,
\label{eqn19}
\end{equation}
where $\bar X = X^{-1}$ and $\bar Y = Y^{-1}$, and hence the isometry
\eqref{eqn3} may be written in the form
\begin{equation}
 J = \sum_{jl} \lm_{jl}\ket{c_l} \ot \bar Y\dyad{b_j}{a_j} \bar X
 = \sum_j (\ket{\bt_j}\ot\ket{\gm_j}) \bra{\bar \al_j},
\label{eqn20}
\end{equation}
where 
\begin{equation}
 \ket{\bt_j} = \nu_j \bar Y\ket{b_j},\quad 
\bra{\bar \al_j} = \mu_j\bra{a_j}\bar X,\quad
\ket{\gm_j} = \sum_l (\lm_{jl}/\mu_j\nu_j)\ket{c_l}, 
\label{eqn21}
\end{equation}
where the $\mu_j$ and $\nu_j$ are positive coefficients.
Since the $\{\bra{\bar\al_j}\}$ are linearly-independent there is a dual basis
$\{\ket{\al_j}\}$ of $\HC_a$ such that
\begin{equation}
 \inpd{\bar\al_j}{\al_k} = \dl_{jk},
\label{eqn22}
\end{equation}
and we assume the positive $\mu_j$ coefficients in \eqref{eqn21} are chosen so
that the $\ket{\al_j}$ (not the $\ket{\bar\al_j}$) are normalized, $\inp{\al_j}
= 1$. The positive $\nu_j$ coefficients are chosen so that $\inp{\bt_j} = 1$.
The right side of \eqref{eqn20}, together with \eqref{eqn22} shows that the
isometry $J$ obtained using twisted-diagonal Kraus operators is of the pcubed
form \eqref{eqn3}. That the $\ket{\gm_j}$ in \eqref{eqn21} are normalized
follows from the fact that $J$ is an isometry, though it can also be checked
using the usual closure condition $\sum_l L_l\ad L^{}_l = I_a$ and some
algebra.

Running the preceding argument in reverse shows that as long as the
$\ket{\al_j}$ and the $\ket{\bt_j}$ (or $\ket{\gm_j}$ if one thinks of the
Kraus operators as mapping $\HC_a$ to $\HC_c$) are linearly independent, i.e.,
they form bases of $\HC_a$ and $\HC_b$ with $d_a=d_b$, pcubed channels are
twisted diagonal channels.  However, if the $\ket{\al_j}$ are linearly
independent but the $\ket{\bt_j}$ are not, the corresponding pcubed isometry
is not of the twisted-diagonal form; for example 
\begin{equation}
 J\ket{1} = \ket{11},\quad J\ket{2} = \ket{12},\quad J\ket{3} = \ket{21},
\label{eqn23}
\end{equation}
where the kets $\ket{j}$ are orthonormal. In the notation of \eqref{eqn16},
$\ket{\bt_1} = \ket{\bt_2} = \ket{1}$ and $\ket{\gm_1} = \ket{\gm_3} =
\ket{1}$, so both the $\ket{\bt_j}$ and the $\ket{\gm_j}$ are linearly
dependent, and neither the $a\ra b$ nor the $a\ra c$ channel is twisted
diagonal. When discussing degradable $a\ra b$ pcubed channels we will assume
that both the $\{\ket{\al_j}\}$ and the $\{\ket{\bt_j}\}$ are linearly
independent, and thus this kind of channel is of the twisted-diagonal form.

\subsection{Degradable pcubed channels}
\label{sbct3.3}

\xb
\outl{Pcubed isometry $K:\HC_b\ra \HC_c\ot\HC_d$; $B=C*D$}
\xa

We assume that the isometry is of the form \eqref{eqn16} with both sets
$\{\ket{\al_j}\}$ and $\{\ket{\bt_j}\}$ linearly independent. In this case the
degrading isometry $K$ in \eqref{eqn10} must necessarily carry each
$\ket{\bt_j}$ to a product state $\ket{\gm_j}\ot\ket{\dl_j}$. The reason is
that $\CC([\al_j])$, where we use the abbreviation $[\psi] = \dya{\psi}$, will
be the pure state $[\gm_j]$, and this means that $\DC([\bt_j])$ must also be
the pure state $[\gm_j]$. This is only possible if $K$ maps $\ket{\bt_j}$ onto
a product of pure states, rather than onto some entangled state, which is to
say $\ket{\gm_j}\ot\ket{\dl_j}$ for a suitable choice of $\ket{\dl_j}$ in
$\HC_d$. But as the $\ket{\bt_j}$ are assumed to be linearly independent, $K$
is the pcubed isometry
\begin{equation}
  K\ket{\bt_j} = \ket{\gm_j}\ot\ket{\dl_j},
\label{eqn24}
\end{equation}
and the analog of \eqref{eqn17} is
\begin{equation}
 B_{jk} = C_{jk} D_{jk}, \text{ or } B = C*D,
 \text{ with } D_{jk} = \inpd{\dl_j}{\dl_k}.
\label{eqn25}
\end{equation}

\xb
\outl{Construct degradable channels using $C$, $D$, or $B$, $C$ $\ra$ $D=C/*B$}
\xa

This provides an easy prescription for constructing a large family of
degradable channels: Choose any two positive definite $d \tm d$ matrices $C$
and $D$, and use \eqref{eqn25} and \eqref{eqn17} to define $B$ and then $A$. The
$a\ra b$ channel with superoperator $\BC$, \eqref{eqn3} will then be a
degradable channel. Or one can choose a positive definite $B$, and $C$ positive
(semi)definite such that their Hadamard product $A$ is positive definite. Then
if $D=C/{*}B$, to use a fairly obvious notation for element-wise division, is
positive semidefinite the $a\ra b$ channel will be degradable, and otherwise
it will not be degradable. Checking the positivity of $D$ is equivalent to
showing that the degrading superoperator $\DC$ is CPTP.  The $H$ matrix whose
positivity was used as a condition for degradability in \cite{WlPG07}
is the same as our $D$ matrix, though showing the equivalence involves some
algebra; thus our condition is equivalent to theirs. 

A large number of examples of channels known to be degradable either belong to
the pcubed family, or are subchannels of degradable pcubed channels, or can be
obtained as limits of pcubed channels by varying a parameter. Some specific
cases are discussed below in Sec.~\ref{sct4}. However, it seems unlikely that
all degradable channels belong to the pcubed family even when that is extended
by the two methods just discussed.

\subsection{Conjugate degradable pcubed channels}
\label{sbct3.4} 

We look for a channel $\BC$ generated by a pcubed isometry \eqref{eqn16},
assuming that both $\{\ket{\al_j}\}$ and $\{\ket{\bt_j}\}$ are linearly
independent (with no similar requirement on the $\{\ket{\gm_j}\}$), such that
there exists a CPTP map $\bar\DC$ satisfying \eqref{eqn13}. We begin by noting
that for the channel pair $\BC$, $\CC$ as defined by inserting \eqref{eqn16} in
\eqref{eqn2},
\begin{equation}
 \BC(\dyad{\al_j}{\al_k}) = C_{kj} \dyad{\bt_j}{\bt_k},\quad
 \CC(\dyad{\al_j}{\al_k}) = B_{kj} \dyad{\gm_j}{\gm_k},
\label{eqn26}
\end{equation}
and, in addition,
\begin{equation}
 \TC(\dyad{\gm_j}{\gm_k}) = \dyad{\bar\gm_k}{\bar\gm_j},
\label{eqn27}
\end{equation}
where the $\{\ket{\bar\gm_j\}}$ are defined by their coefficients
\begin{equation}
 \inpd{c_l}{\bar\gm_j} = \inpd{\gm_j}{c_l} = \inpd{c_l}{\gm_j}^*,
\label{eqn28}
\end{equation}
in the orthonormal basis $\{\ket{c_l}\}$ used in \eqref{eqn14} to define $\TC$.
Note that $j$ and $k$ are in the \emph{opposite order} on the two sides of
\eqref{eqn27}.

Since the dyads $\dyad{\al_j}{\al_k}$ span the operator space $\hat\HC_a$,
${\TC\circ\CC = \bar\DC \circ \BC}$, \eqref{eqn13}, is equivalent to the
requirement that
\begin{equation}
 B_{kj} \dyad{\bar\gm_k}{\bar\gm_j} = C_{kj} \bar\DC(\dyad{\bt_j}{\bt_k})
\label{eqn29}
\end{equation}
be satisfied for every $j$ and $k$.  For $j=k$ (recall that
$B_{jj} = 1 = C_{jj}$) this means that 
\begin{equation}
 \bar\DC([\bt_j]) = [\bar\gm_j],
\label{eqn30}
\end{equation}
where we remind the reader that $[\psi]$ is our notation for the projector
$\dya{\psi}$. Thus, using the same argument as in Sec.~\ref{sbct3.3}, we
see that $\bar \DC$ must be a channel generated by a pcubed isometry of the
form
\begin{equation}
 \bar K\ket{\bt_j} = \ket{\bar\gm_j}\ot \ket{\dl_j},
\label{eqn31}
\end{equation}
and hence
\begin{equation}
 \bar\DC(\dyad{\bt_j}{\bt_k}) = D_{kj} \dyad{\bar\gm_j}{\bar\gm_k},\quad
D_{kj} := \inpd{\dl_k}{\dl_j}.
\label{eqn32}
\end{equation}
Inserting \eqref{eqn32} in \eqref{eqn29} yields
\begin{equation}
 B_{kj} \dyad{\bar\gm_k}{\bar\gm_j} = C_{kj} D_{kj} \dyad{\bar\gm_j}{\bar\gm_k},
\label{eqn33}
\end{equation}
or, upon applying $\TC$ to both sides (note that $\TC\circ\TC$
is the identity) 
\begin{equation}
 B_{kj} \dyad{\gm_j}{\gm_k} = C_{kj} D_{kj} \dyad{\gm_k}{\gm_j}.
\label{eqn34}
\end{equation}

For $j=k$ \eqref{eqn34} is automatically satisfied, since the $B$, $C$, and $D$ 
matrices have $1$'s on the diagonals, while for $j\neq k$ there are
two distinct situations:

Case I: $|\inpd{\gm_j}{\gm_k}| = 1$, so $\ket{\gm_j}$ and $\ket{\gm_k}$ are
identical up to a phase. Consequently, \eqref{eqn34} will be satisfied
provided (take the trace of both sides) 
\begin{equation}
 B_{kj} = C_{jk}D_{kj}.
\label{eqn35}
\end{equation}

Case II. $|\inpd{\gm_j}{\gm_k}| < 1$, so $\ket{\gm_j}$ and $\ket{\gm_k}$ are
not multiples of each other, which means the operators $\dyad{\gm_k}{\gm_j}$
and $\dyad{\gm_j}{\gm_k}$, which do not commute, are also not multiples of each
other. Hence the coefficients on both sides of \eqref{eqn34} must vanish:
\begin{equation}
  B_{kj} = 0 = C_{kj} D_{kj}.
\label{eqn36}
\end{equation}

Both cases may be present for different $j\neq k$ pairs. Hence it is convenient
to segregate the indices $j$ into collections $\JC_m$, such that $j$ and $k$
are in the same collection if $\ket{\gm_j}$ and $\ket{\gm_k}$ are equal up to a
phase, Case I, and $j$ and $k$ belong to different collections when
$\ket{\gm_j}$ and $\ket{\gm_k}$ are not multiples of each other, Case II. Let
us add a superscript to each ket to indicate the corresponding collection:
$\ket{\al_j^m}=\ket{\al_j}$, $\ket{\bt_j^m}=\ket{\bt_j}$, etc., when $j$ is a
member of $\JC_m$. With this notation the isometry $J$, \eqref{eqn16}, can be
written as
\begin{equation}
  J \ket{\al_j^m} = \ket{\bt_j^m}\ot\ket{\gm^m},
\label{eqn37}
\end{equation}
where $\ket{\gm^m}$ is $\ket{\gm_j^m}$ for some $j\in \JC_m$, and the phases of
the others have been absorbed in the definitions of the $\ket{\bt_j^m}$, which
involves no loss of generality.
Then \eqref{eqn26} can be rewritten in the form:
\begin{equation}
 \BC(\dyad{\al_j^m}{\al_k^n}) = C^{nm} \dyad{\bt_j^m}{\bt_k^n},\hquad
C^{nm} := \inpd{\gm^n}{\gm^m}; \quad
\CC(\dyad{\al_j^m}{\al_k^n}) = B_{kj} \dl_{mn} [\gm^m],
\label{eqn38}
\end{equation}
where for $m\neq n$, \eqref{eqn37} implies that
$B_{kj}=\inpd{\bt_c^n}{\bt_j^m}=0$.  The isometry $\bar K$ in \eqref{eqn31}
has the form:
\begin{equation}
 \bar K\ket{\bt_j^m} = \ket{\bar\gm^m}\ot \ket{\dl_j^m},
\label{eqn39}
\end{equation}
and generates the channel $\bar D$, \eqref{eqn32} required to show that 
$\BC$ is conjugate degradable.

However, if we use the isometry
\begin{equation}
 K\ket{\bt_j^m} = \ket{\gm^m}\ot \ket{\dl_j^m},
\label{eqn40}
\end{equation}
in which $\ket{\bar\gm^m}$ in \eqref{eqn39} has been replaced by $\ket{\gm^m}$,
the corresponding $\DC$, \eqref{eqn11}, satisfies $\CC = \DC\circ\BC$,
\eqref{eqn12}, and hence $\BC$ is both degradable and conjugate degradable.
The preceding argument employs special features of a pcubed approach, so it
leaves open the possibility that there may exist conjugate degradable channels
that are \emph{not} degradable.  But perhaps it provides some insight into which
channels one should study.

\section{Examples}
\label{sct4}

\subsection{Qubit to qubits}
\label{sbct4.1}

For $d_a=d_b=d_c=2$ both channels of the complementary pair map a qubit input
to a qubit output. It is known that all such pairs are degradable
\cite{WlPG07}, so any channel of this type is either degradable or
antidegradable (or both). The pcubed approach leads to the same result from a
slightly different perspective, so going through it is a useful exercise.
Instead of 1 and 2 we use 0 and 1 to label the kets in \eqref{eqn16}, and the
rows and columns of the corresponding matrices. The four $2\tm 2$ matrices $A$,
$B$, $C$, and $D$ have 1's on the diagonal, and the two off-diagonal elements
are complex conjugates of each other. Equations \eqref{eqn17} and \eqref{eqn25}
will be satisfied provided
\begin{equation}
 A_{01}  = B_{01}C_{01},\quad B_{01}=C_{01}D_{01}
\label{eqn41}
\end{equation}
Each matrix is positive (semi)definite if the magnitude of the off-diagonal
element is less than (or equal to) 1. Thus the condition that the $a\ra b$
channel with superoperator $\BC$ be degradable corresponds to the requirement
that $|B_{01}| \leq |C_{01}| \leq 1$.

A degradable channel pair of this type has a simple geometrical interpretation.
The map $\BC$ carries density operators forming a Bloch sphere in the channel
input $a$, into an ellipsoid lying inside the Bloch sphere of $b$, but touching
it at the two points $[\bt_0] = \dya{\bt_0}$ and $[\bt_1] = \dya{\bt_1}$
corresponding to pure states. The fact that $|A_{01}|$ is less than $|B_{01}|$
means that these points are closer together, less distinguishable, than the
corresponding pure states $[\al_0]$ and $[\al_1]$ at the channel entrance. The
channel is degradable when $|B_{01}| \leq |C_{01}|$, so the distance on the
Bloch sphere between $[\bt_0]$ and $[\bt_1]$ is greater than that between
$[\gm_0]$ and $[\gm_1]$ for the complementary channel. When $|B_{01}| \geq
|C_{01}|$ the channel $\CC$ is degradable and $\BC$ is antidegradable, as the
distance between $[\gm_0]$ and $[\gm_1]$ is greater than that between $[\bt_0]$
and $[\bt_1]$, while for $|B_{01}| = |C_{01}|$ $\BC$ and $\CC$ are both
degradable and antidegradable.

It is convenient to write
\begin{alignat}{2}
 \ket{\al_0} &= a_0\ket{0} + a_1\ket{1},&\quad 
\ket{\al_1} &= a_0\ket{0} - a_1\ket{1},
\notag\\
 \ket{\bt_0} &= b_0\ket{0} + b_1\ket{1},&\quad 
\ket{\bt_1} &= b_0\ket{0} - b_1\ket{1},
\notag\\
 \ket{\gm_0} &= c_0\ket{0} + c_1\ket{1},&\quad
\ket{\gm_1} &= c_0\ket{0} - c_1\ket{1}
\label{eqn42}
\end{alignat}
using the standard orthonormal basis $\ket{0}$ and $\ket{1}$, with
$a_0,\,a_1,\,b_0$, etc.  real numbers between 0 and 1 chosen so that the
kets are normalized: $a_0^2+a_1^2 = 1$, etc. The off-diagonal matrix element
$A_{01}=A_{10}$ is $a_0^2-a_1^2$, and $B_{01}$ and $C_{01}$ are given by
analogous expressions. The condition $|B_{01}| \leq |C_{01}|$ for the $a\ra b$
channel to be degradable is equivalent to $b_1 \geq c_1$.

The isometry $J\ket{\al_j} = \ket{\bt_j}\ot \ket{\gm_j}$ using the definitions
in \eqref{eqn42} takes the form
\begin{equation}
 J\ket{0} = f_0\ket{00} + f_1\ket{11},\quad
 J\ket{1} = g_0\ket{01} + g_1\ket{10},\quad
\label{eqn43}
\end{equation}
in the standard basis, where again the $f_j$ and $g_j$ are real numbers
between 0 and 1, $f_0^2+f_1^2 = g_0^2 + g_1^2 = 1$, and
\begin{equation}
 f_0 = b_0 c_0/a_0,\quad f_1 = b_1 c_1/a_0,\quad
 g_0 = b_0 c_1/a_1,\quad g_1 = b_1 c_0/a_1.
\label{eqn44}
\end{equation}
expresses them in terms of the parameters in \eqref{eqn42}. 

Any qubit to qubits isometry can be written, up to local unitaries, in the form
\eqref{eqn43} with a suitable choice of the $f_j$ and $g_j$,
since a two-dimensional subspace of two qubits always possesses an orthonormal
basis of the form shown on the right side \cite{NiGr99}. Therefore such an
isometry is of the pcubed form provided there are values of $a_j$, $b_j$, and
$c_j$ such that \eqref{eqn44} is satisfied. This is easily seen to be the case
unless one of the $f_j$ or $g_j$ is 0. If, for example $f_1=0$, either $b_1$ or
$c_1$ must be 0, so either $g_0$ or $g_1$ will vanish. Hence the isometry
\eqref{eqn43} with
\begin{equation}
 f_0 = 1,\quad f_1=0,\quad g_0 = \sqrt{p},\quad g_1 = \sqrt{1-p},
\label{eqn45}
\end{equation}
with $0<p<1$, so both $g_0$ and $g_1$ are positive, corresponding to the
amplitude damping channel, illustrates the exceptional situation in which
\eqref{eqn43} is \emph{not} pcubed.
From a geometrical point of view this situation corresponds to a limiting case
in which the contact points between the ellipsoid and the Bloch sphere, for
both the $\BC$ and $\CC$ channels, coalesce into a single point, which suggests
that the isometry corresponding to \eqref{eqn45} is the limit of pcubed
isometries. To see that this is indeed the case, let $a_1$, $b_1$, and $c_1$ be
small positive quantities chosen so that
\begin{equation}
 c_1/a_1 = \sqrt{p},\quad b_1/a_1 = \sqrt{1-p},
\label{eqn46}
\end{equation}
and take the limit as $a_1$ tends to zero, with $p$ fixed. The result is that
the $f_j$ and $g_j$ parameters tend continuously to those in \eqref{eqn45}.
The condition for the $a\ra b$ channel $\BC$ to be degradable, $b_1 \geq c_1$,
is fulfilled for $p\leq 1/2$, and one can check that the pcubed isometry
\eqref{eqn24} for the degrading map exists and tends to the appropriate limit
as long as $p\leq 1/2$. Thus all qubit-to-qubits channel pairs are given either
directly or as a limit of pcubed isometries whose degradability is determined
by whether the $D=B/*C$ matrix, \eqref{eqn25} is positive (semi)definite.

\subsection{Erasure channel}
\label{sbct4.3}

A one qubit erasure channel $a\ra b$ transmits a linear combination of
$\ket{0}$ and $\ket{1}$ unchanged to the output space $\HC_b$ with a
probability $1-p$, but with probability $p$ replaces it with the error flag
$\ket{e}$. It can be represented by the $\HC_a$ to $\HC_b\ot\HC_c$ isometry
\begin{equation}
 J\ket{0} = \sqrt{1-p}\,\ket{0e} + \sqrt{p}\,\ket{e0},\quad
 J\ket{1} = \sqrt{1-q}\,\ket{1e} + \sqrt{1-q}\,\ket{e1},\quad
 J\ket{e} = \ket{ee}
\label{eqn47}
\end{equation}
where we have added a state $\ket{e}$ to the input space so that
$d_a=d_b=d_c=3$, and introduced an additional probability $q$ which when 
set equal to $p$ yields the usual erasure channel. The kets $\ket{0}$,
$\ket{1}$, and $\ket{e}$ form an orthonormal basis for each of the Hilbert
spaces. The usual erasure channel (for $q=p$) with input the subspace of
$\HC_a$ spanned by $\ket{0}$ and $\ket{1}$ is a subchannel of the channel
defined by \eqref{eqn47},

The isometry \eqref{eqn47} is the limit of a pcubed isometry of the form
\eqref{eqn16} in which
\begin{alignat}{3}
 \ket{\bt_0} &= (\ket{e} + \zt\sqrt{1-p}\,\ket{0})/b_0,&\quad
 \ket{\bt_1} &= (\ket{e} + \eta\sqrt{1-q}\,\ket{1})/b_1,&\quad
 \ket{\bt_e} &= \ket{e},
 \notag\\
 \ket{\gm_0} &= (\ket{e} + \zt\sqrt{p}\,\ket{0})/c_0,&\quad
 \ket{\gm_1} &= (\ket{e} + \eta\sqrt{q}\,\ket{1})/c_1,&\quad 
 \ket{\gm_e} &= \ket{e},
\label{eqn48}
\end{alignat}
where the normalization factors are
\begin{equation}
 b_0 = \sqrt{1+(1-p)\zt^2}, \quad 
 b_1 = \sqrt{1+(1-q)\eta^2}, \quad 
 c_0 = \sqrt{1+p\,\zt^2}, \quad 
 c_1 = \sqrt{1+q\,\eta^2}, \quad 
\label{eqn49}
\end{equation} 
where $\zt$ and $\eta$ are small positive numbers which tend to zero. To see
that the isometry so defined converges to \eqref{eqn47}, note
that the subspace of $\HC_b\ot\HC_c$ that contains $\ket{\bt_0}\ot\ket{\gm_0}$,
$\ket{\bt_1}\ot\ket{\gm_1}$, and $\ket{\bt_e}\ot\ket{\gm_e}$ is spanned by
$\ket{ee}$ and two other kets which differ from their counterparts on the right
side of \eqref{eqn47} by small corrections of order $\zt$ and $\eta$.

The $3\tm 3$ $B$ and $C$ matrices constructed from \eqref{eqn48} (see
\eqref{eqn17}) have $1$'s on the diagonal and off-diagonal elements (note that
$B_{e0}=B_{0e}$, etc.)
\begin{equation}
 B_{0e} = 1/b_0,\quad B_{1e} = 1/b_1,\quad B_{01} = 1/b_0b_1,\quad
 C_{0e} = 1/c_0,\quad C_{1e} = 1/c_1,\quad C_{01} = 1/c_0c_1.
\label{eqn50}
\end{equation}
As these are real and positive, the same will be true of the off-diagonal
elements of the matrix $D = B/*C$, \eqref{eqn27}, so $D$ will satisfy the
positivity conditions given in \eqref{eqn48}, and the full $a\ra b$ channel
defined by \eqref{eqn47}, thus also the subchannel with inputs spanned by
$\ket{0}$ and $\ket{1}$, will be degradable, as long as $B_{jk}\leq C_{jk}$, or
\begin{equation}
 0 \leq p \leq 1/2, \quad 0 \leq q \leq 1/2.
\label{eqn51}
\end{equation}

The same argument works for a $d$-dimensional erasure channel with input space
spanned by $\ket{0},\,\ket{1},\ldots \ket{{d-1}}$ and erasure probabilities
$p_0, p_1, \ldots$ playing the role of $p$ and $q$ in \eqref{eqn47}. By making
this a subchannel of the $d+1$-dimensional channel obtained by adding $\ket{e}$
to the input space and using the obvious analogs of \eqref{eqn48} and can
demonstrate degradability provided all the $p_j$ are no greater than 1/2.

\subsection{Entanglement breaking and Hadamard}
\label{sbct4.4}

An entanglement-breaking channel \cite{HrSR03} can be defined in various ways,
of which the most convenient for our purposes is that it is generated by Kraus
operators of rank 1. The complement of an entanglement-breaking channel is a
Hadamard channel, and is known to be degradable \cite{KMNR07}. We will show that
a Hadamard channel is a subchannel of a pcubed channel.

If the $a\ra c$ channel is entanglement breaking, then
it and the complementary Hadamard channel are generated by an isometry of the
form
\begin{equation}
 \hat J = \sum_j \ket{b_j}\ot \ket{\gm_j}\bra{\hat\al_j},
\label{eqn52}
\end{equation}
where the $\ket{b_j}$ form an orthonormal basis of $\HC_b$ and the $L_j=
\ket{\gm_j}\bra{\hat\al_j}$ are Kraus operators mapping $\HC_a$ to $\HC_c$.
There is no assumption that the collections $\ket{\gm_j}$ and $\ket{\hat\al_j}$
are orthogonal. For convenience we assume the $\ket{\gm_j}$ are
normalized (as can always be arranged using a suitable normalization for the
$\ket{\hat\al_j}$), which allows the completeness condition for the Kraus
operators to be written in the form
\begin{equation}
 \sum_j L_j\ad L^{}_j = \sum_j\dya{\hat\al_j} = I_a.
\label{eqn53}
\end{equation}
Thus the $d_b$ positive operators $\dya{\hat\al_j}$ form a POVM. If $d_b$ is
larger than $d_a$ the Naimark extension (or dilation) theorem \cite{Jzao03}
tells us there exists a Hilbert space $\HC_A$ with an orthonormal basis
$\ket{a_j}$ which contains $\HC_a$ as a subspace, such that
\begin{equation}
 P\ket{a_j} = \ket{\al_j},
\label{eqn54}
\end{equation}
where $P$ is the projector in $\hat\HC_A$ which projects onto $\HC_a$ regarded
as a subspace of $\HC_A$. (If $d_b=d_a$ then $\HC_A$ is the same as $\HC_a$,
and $\ket{\al_j} = \ket{\hat\al_j}$.)
We now define an isometry $J:\HC_A \ra \HC_b\ot\HC_c$ by the formula
\begin{equation}
 J\ket{a_j} = \ket{b_j}\ot\ket{\gm_j},
\label{eqn55}
\end{equation}
which is \eqref{eqn16} with $\ket{\al_j} = \ket{a_j}$ and $\ket{\bt_j} =
\ket{b_j}$, and thus pcubed. That it is an isometry follows from the fact that
the $\ket{a_j}$ and $\ket{b_j}$ are orthonormal bases, and the $\ket{\gm_j}$
are normalized. Thus $A$ and $B$ in \eqref{eqn17} are the identity matrices, and
making $D$ the identity matrix means \eqref{eqn54} will also be satisfied.
Hence $J$ defines a degradable $A\ra b$ channel. But since any subchannel of a
degradable channel, obtained by restricting the channel entrance to a subspace
while leaving the exit spaces $\HC_b$ and $\HC_c$ unchanged, is (obviously)
also degradable, the Hadamard channel induced by $\hat J = JP$ is also
degradable.

\section{Qutrit to qutrits}
\label{sct5}

In the case of qutrits, $d_a = d_b = d_c=3$, the $3\times 3$ matrices
of Sec.~\ref{sbct3.1} are of the form
\begin{equation}
M = \mat{ 1 & m_1    & m_2 \\
        m_1^* & 1    & m_3 \\ 
        m_2^* & m_3^*& 1},
\label{eqn56}
\end{equation}
where the $m_j$ are complex numbers, equal to $a_j$, $b_j$, and $c_j$ for the
matrices $A$, $B$ and $C$, with $a_j=b_j c_j$.   The necessary and sufficient conditions
for $M$ to be positive semidefinite are:
\begin{equation}
 |m_1| \leq 1,\hquad |m_2| \leq 1,\hquad |m_3| \leq 1,\hquad 
 |m_1|^2 + |m_2|^2 + |m_3|^2 \leq 1 + 2\, \Re(m^{}_1 m_2^* m^{}_3),
\label{eqn57}
\end{equation}
and it will be positive definite if all the inequalities are strict.

\subsection{Equal off-diagonal elements}
\label{sbct5.1}

A very simple situation is that in which the off-diagonal elements 
are equal real numbers
\begin{equation}
 b_j = b,\hquad c_j = c,\hquad a_j = a = bc.
\label{eqn58}
\end{equation}
The inequalities \eqref{eqn57} will be satisfied provided $-1/2 \leq b,c \leq
1$, which ensures that $a=bc$ falls in the same range. 
\ca
As we shall see, even
this simple example exhibits interesting features that goes beyond anything
observed in the qubit to qubits case discussed in Sec.~\ref{sbct4.1}.
\cb
The $\BC$ channel is degradable and the $\CC$ channel anti-degradable when $D =
B/*C$, \eqref{eqn25} is positive semi-definite, which is to say,
\begin{equation}
-1/2 \leq b/c \leq 1, 
\label{eqn59}
\end{equation}
whereas $\CC$ is degradable and $\BC$
anti-degradable for $c/b$ between $-1/2$ and 1.

\begin{figure}[p]
\begin{center}
\includegraphics[scale = .67]{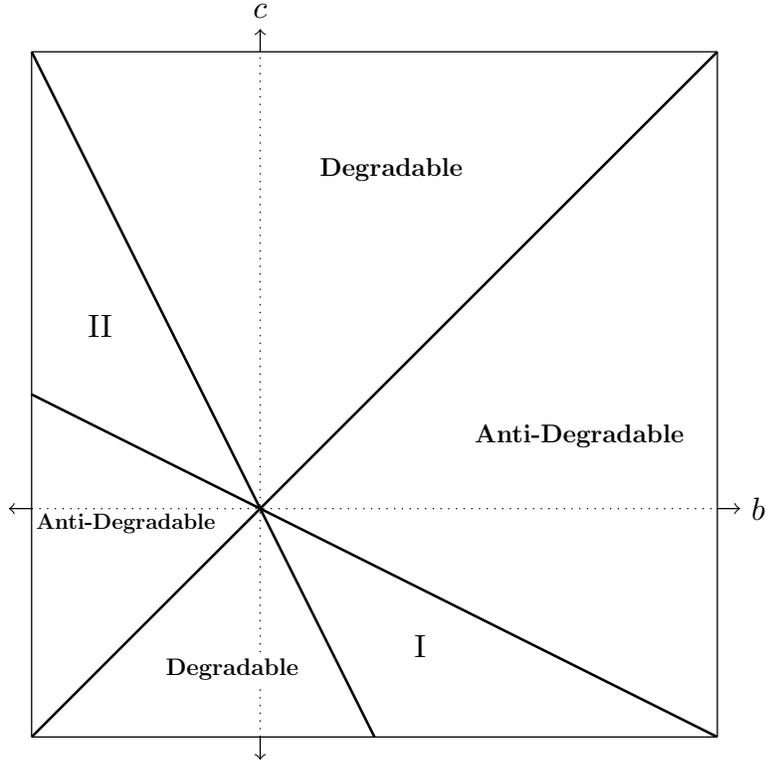}
\caption{The $(b,c)$ plane phase diagram with $b$ and $c$ between $-1/2$ and $1$.
The $\BC$ channel is degradable ($\CC$ anti-degradable) or anti-degradable
($\CC$ degradable) in the wedges carrying these labels, while in  I and II
the channel pair is undegradable (neither $\BC$ nor $\CC$ is either degradable
or antidegradable). \label{fgr1}}
\end{center}
\end{figure}

\begin{figure}[]
\begin{center}
\includegraphics[scale = 1.15]{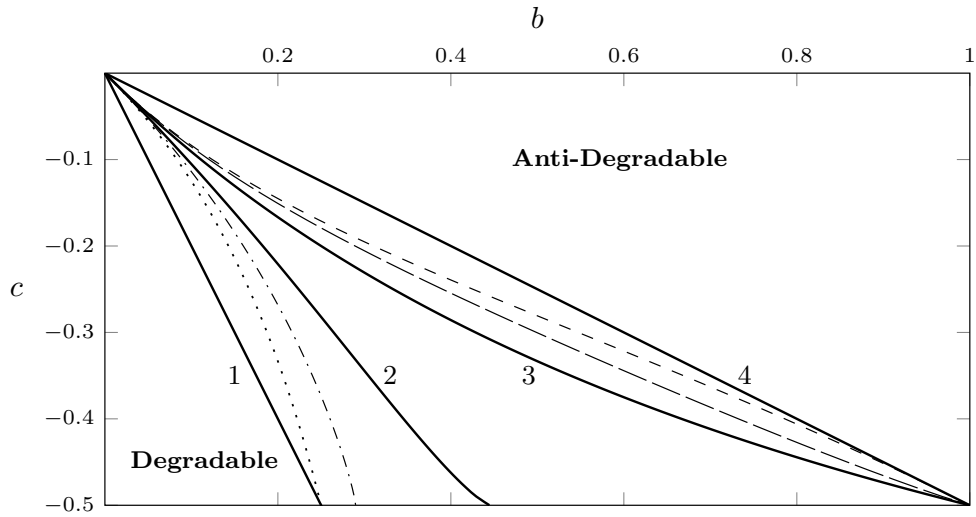}
  \caption{The region $b\geq 0$ and $c\leq 0$ showing details of region I in
    Fig.~\ref{fgr1}, which lies between the straight lines 1 and 4. See text
    for the significance of the different curves.
\label{fgr2}}
\end{center}
\end{figure}

Figure~\ref{fgr1} shows the regions in the $b,c$ plane where $\BC$ is
degradable ($\CC$ anti-degradable) or anti-degradable ($\CC$ degradable), which
are mapped into each other by symmetry when $b$ and $c$ are interchanged. This
symmetry also implies that both one-shot capacities $Q^{(1)}(\BC)$ and
$Q^{(1)}(\CC)$, \eqref{eqn6} and \eqref{eqn7}, vanish along the line $b=c$. In
addition, in the regions labeled I and II the $\BC$ and $\CC$ channels are
\emph{undegradable} as defined in Sec.~\ref{sbct2.3}: each is
neither degradable nor is it antidegrable: neither $b/c$ nor $c/b$ lies
between $-1/2$ and 1. 

Figure~\ref{fgr2} shows the $b\geq 0$ and $c\leq 0$ rectangle that contains
wedge I of Fig.~\ref{fgr1}, with additional information from numerical studies
of the input density operators $\rho$ on $\HC_a$ which maximize and minimize
$\Dl S(\rho)$, \eqref{eqn5}, and thus determine $Q^{(1)}(\BC)$ and
$Q^{(1)}(\CC)$. The one shot capacity $Q^{(1)}(\BC)$ is positive everywhere to
the left of curve 3, where it goes continuously to zero, and remains zero for
larger $b$. Similarly, $Q^{(1)}(\CC)$ is positive everywhere to the right of
curve 2 where it rises continuously from zero, and zero everywhere to the left
of this curve. Hence there is a region between curves 2 and 3 where
$Q^{(1)}(\BC)$ and $Q^{(1)}(\CC)$ are positive, something impossible for a
degradable channel pair. The other curves in the figure indicate boundaries of
concavity or convexity of $\Dl S(\rho)$, \eqref{eqn5}, and points at which the
optimizing density operator changes its character, as discussed next.

\subsection{Optimizing density operators \label{sbct5.2}}

Understanding what happens in the wedge between lines 1 and 4 in
Fig.~\ref{fgr2} (region I in Fig.~\ref{fgr1}) requires paying some attention to
the symmetries of the optimizing density operators.
The fact that the off-diagonal elements of the $A$ matrix are equal makes it
invariant under any permutation of the three kets $\ket{\al_j}$ for $j=1,2,3$,
the symmetry group $S_3$, or $D_2$ for an equilateral triangle, and likewise
$B$ and $C$ upon permuting the $\ket{\bt_j}$ and of the $\ket{\gm_j}$. A
convenient orthonormal or \emph{standard} basis $\{\ket{0}, \ket{1}, \ket{2}
\}$ of $\HC_a$ is one in which the nonorthogonal kets $\ket{\al_j}$ take the
form:
\begin{alignat}{2}
 \ket{\al_1} &= \sqrt{\frac{1 + 2a}{3}} \ket{0} - \sqrt{\frac{1 - a}{2}} \ket{1}
 -\sqrt{\frac{1 - a}{6}} \ket{2},
\notag\\
 \ket{\al_2} &= \sqrt{\frac{1 + 2a}{3}} \ket{0} + \sqrt{\frac{1 - a}{2}} \ket{1}
 -\sqrt{\frac{1 - a}{6}} \ket{2},
\notag\\
 \ket{\al_3} &= \sqrt{\frac{1 + 2a}{3}} \ket{0} + 
\sqrt{\frac{2(1 - a)}{3}} \ket{2},
\label{eqn60}
\end{alignat}
where on can visualize the components of the $\ket{\al_j}$ in the
$\ket{1},\,\ket{2}$ plane as vertices of an equilateral triangle. We use
similar standard bases of $\HC_b$ and $\HC_c$ in which the $\{\ket{\bt_j}\}$
and $\{\ket{\gm_j}\}$ have the form shown in \eqref{eqn60}, but with $a$
replaced by $b$ and by $c$.

A representation of the $D_2$ ($S_3$) symmetry is provided by powers and
products of the orthogonal (hence unitary) matrices
\begin{equation}
F = \mat{1 & 0 & 0\\
0 & -1 & 0\\
0 & 0 & 1},
\quad 
R = \mat{1 & 0 & 0 \\
0 & -1/2 & -\sqrt{3}/2 \\
0 & \sqrt{3}/2 & -1/2},
\label{eqn61}
\end{equation}
using the standard basis in \eqref{eqn60}, or its counterparts for
$\HC_b$ and $\HC_c$. Here $F$ is a ``reflection'' that interchanges
$\ket{\al_1}$ and $\ket{\al_2}$, whereas $R$ is a ``rotation'' $\ket{\al_1} \ra
\ket{\al_2}\ra \ket{\al_3}\ra \ket{\al_1}$. The pcubed isometry $J$ in
\eqref{eqn16} is symmetrical in the sense that
\begin{equation}
 J G_a = (G_b \ot G_c) J,
\label{eqn62}
\end{equation}
where $G$ is any one of the matrices of the group representation, and the
subscript indicates which space, $\HC_a$ or $\HC_b$ or $\HC_c$, it is acting
on. In light of \eqref{eqn62} the superoperators defined in \eqref{eqn2} have
the property that
\begin{equation}
 \BC(G_a^{}\, \rho\, G_a^\dag) =  G_b^{}\, \BC(\rho)\, G_b^\dag, \quad
\CC(G_a^{}\, \rho\, G_a^\dag) =  G_c^{}\, \CC(\rho)\, G_c^\dag,
\label{eqn63}
\end{equation}
where $\rho$ is an input density operator. In addition the fact that the
parameters $b$ and $c$ are real numbers means that the full symmetry group for
the superoperators includes complex conjugation in the standard basis, an
antiunitary operation that commutes with the other group elements; for this
case think of $G\rho\, G^\dag$ in \eqref{eqn63} as the complex conjugate (or
transpose) of the density operator in the standard basis.

Our numerical studies indicate that the optimizing density operator $\rho$ on
$\HC_a$ that yields the maximum, $Q^{(1)}(\BC)$, or minimum, $-Q^{(1)}(\CC)$,
of $\Dl S(\rho)$ (see \eqref{eqn5}) is always of the form
\begin{equation}
\rho_R = 
 \mat{ 1 - 2s & 0    & 0 \\
      0 & s    & it \\ 
        0 & -it & s},
\label{eqn64}
\end{equation}
for some choice of the two real parameters $0\leq s\leq 1/2$ and $0\leq t\leq
s$, where the subscript $R$ indicates that $\rho_R$ is invariant under the
(``rotation'') subgroup generated by the matrix $R$ in \eqref{eqn61}. Indeed, we
find that the optimum input $\rho$ is always one of three special cases:
\begin{equation}
 \rho_0 = \mat{
 1-2s & 0 & 0\\
 0    & s & 0\\
 0    & 0 & s},\quad
 \rho_1 = \mat{
 0  & 0   & 0\\
 0  & 1/2 & 0\\
 0  & 0   & 1/2},\quad
 \rho_2 = \mat{
 1-2s & 0    & 0\\
 0    & s    & i s\\
 0    & -i s & s }.
\label{eqn65}
\end{equation}
obtained from $\rho_R$ by setting $t=0$, or $t=0$ and $s=1/2$,
or $t=s$. Note that both $\rho_1$ and $\rho_2$ are rank 2 (one eigenvalue equal
to zero), and their supports are two-dimensional subspaces that do not
intersect (except for the origin) for $s< 1/2$.  
The diagonal $\rho_0$ is the most general matrix invariant under the full
symmetry group, which means that its images under $\BC$ and $\CC$ will also
have this form (but of course with different values of the parameter s). The
images of $\rho_1$ will be of the form $\rho_0$, but with $s$ not equal to
$1/2$ in general, while those of $\rho_2$ will be of the form $\rho_R$, with,
in general, $t < s$ rather than $t=s$.

For values of $(b,c)$ falling within the degradable (or antidegradable)
regions in Fig.~\ref{fgr1} the input density operator that maximizes (or
minimizes) $\Dl S(\rho)$ is always of the form $\rho_0$, a plausible
consequence of the fact that degradability implies the concavity of $\Dl
S(\rho)$ as a function of $\rho$, so one expects a unique maximum. Since
when a maximum is achieved at some $\rho$, it is also, by symmetry, achieved at
$G_a\rho G_a^\dag$, these two must be identical if the maximum is unique. When
$\CC$ is degradable and $\BC$ antidegradable the same argument applies to the
minimum of the convex function $\Dl S(\rho)$.

In the region between lines 1 and 4 in Fig.~\ref{fgr2} where the channel pair
is undegradable, there is no general reason to expect that $\Dl S(\rho)$ will
be either concave or convex. Our numerical studies indicate that $\Dl S(\rho)$,
which is necessarily concave in the degradable region to the left of line 1,
continues to be concave up to the dotted line. Here the breakdown of concavity
begins at the pure state $\rho_0$ with $s=0$ and, as $b$ increases, manifests
itself in the fact that the second derivative of $\Dl S(\rho_R)$ with respect
to $t$ evaluated at $t=0$ is positive (contrary to concavity) for $s$ lying in
an interval $(0,s_0)$. As $b$ increases $s_0$ increases until it arrives at the
value of $s$ that maximizes $\Dl S(\rho_0)$, which occurs at the dot-dash line
in the figure. At this point the optimizing density operator changes
discontinuously from $\rho_0$ to $\rho_2$ with increasing $b$, though
$Q^{(1)}(\BC)$ is continuous. For larger values of $b$ the optimizing density
operator continues to have the form $\rho_2$, with $s$ depending on $b$ and $c$,
until $Q^{(1)}(\BC)$ goes to zero continuously at line 3.

Analogous changes occur as $b$ decreases across line 4 in Fig.~\ref{fgr2}, the
boundary of the degradable region for $\CC$, where $\Dl S(\rho)$ must be
convex. Our numerical studies indicate that convexity continues to hold to the
right of the dashed line with smaller dashes just to the left of line 4, and
it begins to break down at the pure state $\rho_0$ with $s=0$, but this
time the signature is a change in the sign of $\pr^2 \Dl S/\pr s^2$. As $b$
continues to decrease the value of $s$ that minimizes $\Dl S$ increases until
it reaches its maximum value of $1/2$, at the dashed line with the larger
dashes. At still smaller values of $b$ the optimizing (minimizing) density
operator is $\rho_1$, all the way until $Q^{(1)}(\CC)$ goes to zero at curve 2.  

Some additional details including formulas for the curves in Fig.~\ref{fgr2}
are given in the Appendix.

\subsection{Summary \label{sbct5.3}}

Let us summarize what seem to be the most significant features found in the
wedge region between solid lines 1 and 4 in Fig.~\ref{fgr2} (I in
Fig.~\ref{fgr1}) where the channel pair is undegradable. First, in the region
between solid lines 2 and 3, both $Q^{(1)}(\BC)$ and $Q^{(1)}(\CC)$, the
one-shot capacities of the two complementary channels, are positive, and in
this overlap region the two optimizing density operators are both of rank 2,
supported on nonintersecting subspaces of $\HC_a$. The regions where $\Dl
S(\rho)$ is either concave or convex extend slightly outside the regions where
these properties are guaranteed by the degradability of $\BC$ and of $\CC$,
respectively, and the breakdown of concavity or convexity begins at a pure
state on the boundary of the convex set of input density operators. There are
in addition some qualitative changes in the character of the optimizing density
operators. For $Q^{(1)}(\BC)$ this is connected to a breakdown of local
concavity of $\Dl S(\rho)$ at its maximum, resulting in a change in the
optimizing density operator from $\rho_0$ to the less symmetric $\rho_2$ at the
dot-dash line. In the case of $Q^{(1)}(\CC)$ it arises when $s$ in $\rho_0$ at
the minimum of $\Dl S(\rho)$ arrives at its maximum value of $1/2$.

\section{Conclusion}
\label{sct6}
 
We have shown that pcubed, ``pure to product of pure states'', isometries are a
useful tool for constructing examples of degradable quantum channels, since
degradability is relatively easy to confirm by checking the positivity of the
$D$ matrix defined in Sec.~\ref{sbct3.3}. As this is a $d \tm d$ matrix,
checking its positivity is generally simpler than that of the corresponding
$d^2 \tm d^2$ Choi matrix of a comparable channel. Furthermore, by simply
choosing positive $C$ and $D$ matrices one obtains, via the Hadamard
(entry-wise) products $B=C*D$ and $A=C*C*D$, an $a\ra b$ channel which is
guaranteed to be degradable, and whose $Q=Q^{(1)}$ quantum capacity is therefore
relatively easy to calculate. Additional examples of degradable channels can be
constructed using subchannels of degradable pcubed channels, or the limit of a
family of such channels depending on a parameter, or a combination of these two
procedures, as illustrated by the examples in Sec.~\ref{sct4}. It is even
conceivable, though it seems unlikely, that all degradable channels can be
obtained from degradable pcubed channels in this manner. In any case the
connection between pcubed and degradability could be a fruitful area for
further study.

The family of pcubed channels includes much more than just degradable channels.
Indeed, since an isometry corresponds (ignoring unitaries on $\HC_a$) to a
particular subspace of $\HC_b\ot \HC_c$, and as it is easy to construct (in
many different ways) a basis of $\HC_b\ot \HC_c$ using product states,
\emph{any} quantum channel is a subchannel of \emph{some} pcubed channel. By
itself this observation is of no great significance. However, as shown by the
specific example in Sec.~\ref{sct5} of a qutrit mapped to a product of qutrits,
the pcubed construction allows the construction of simple examples of channel
pairs which are neither degradable nor antidegradable, but nonetheless can have
nonzero one-shot capacities, and by varying a parameter one can move
continuously from a degradable to a undegradable channel pair and ask what
happens to the one-shot capacity, and to the concavity/convexity properties of
the coherent information as a function of the input density operator. The
interesting behavior in the very simple case of equal off-diagonal
elements of the $A$, $B$, and $C$ matrices, Secs.~\ref{sbct5.1} and
\ref{sbct5.2}, is not something we would have anticipated in advance of
numerical calculations. These features may or may not prove useful in future
studies of quantum capacity and its additivity properties, they at
least seem worth thinking about. 

From a somewhat broader perspective, our work suggests it may be useful to
think about properties of quantum channels in terms of the corresponding
isometries, even when these are not of the pcubed type, rather than placing the
entire emphasis on channel superoperators, as in much of the published
literature. To be sure, there is a one-to-one correspondence, up to local
unitaries, between isometries and completely positive trace preserving maps.
Nevertheless, relationships and intuitive insight can sometimes emerge more
readily from one perspective than from the other, and thus both need to be kept
in mind.

\xb
\section*{Acknowledgments}
\xa

We are indebted to Edward Gerjuoy for numerous discussions and comments about
degradable channels and our pcubed approach. The research described here
was supported by the National Science Foundation (NSF) through Grant 1068331.
Numerical computations at the Pittsburgh Supercomputing Center (PSC) received NSF
support through Grant OCI-1053575 (Extreme Science and Engineering Discovery
Environment or XSEDE) and Award ACI-1261721 for the Greenfield system. We are
also grateful for the help provided by the PSC staff. 

\appendix
\section{Appendix. Details of  Fig.~\ref{fgr2}\label{scta}}

Numerical studies were carried out at several thousands of points in the part
of the $b,c$ plane shown in Fig.~\ref{fgr2} to find the maximum and minimum
values of $\Dl S(\rho)$, setting $\rho =R\ad R$ with $R$ an upper triangular
matrix depending on 3 real parameters for the diagonal elements and 6 real
parameters representing the real and imaginary parts of the off diagonal
elements, with the sum of the squares equal 1. We used standard numerical
optimization techniques (Mathematica). Concavity and convexity were checked by
evaluating the Hessian of $\Dl S(\rho)$ for several thousand randomly-chosen
density operators, at a large number of $b,c$ points.
Once the general character of the optimizing density operators emerged, as
discussed in Sec.~\ref{sbct5.3}, we obtained expressions determining the
different curves in Fig.~\ref{fgr2} as discussed below, and checked that
these were in good agreement with our earlier numerical studies. 

The density operator $\rho_R$ in \eqref{eqn64} depends on two parameters $s$
and $t$, and by symmetry $\BC(\rho_R)$ and $\CC(\rho_R)$ are of the same form
but with different values of the parameters, which we denote by $s_b$, $t_b$
and $s_c$,$t_c$.  Thus 
\begin{equation}
 s_b = \frac{1-b}{1+2bc}\left[ \frac{1-c}{3} + \frac{(1+2b)c}{1-bc} s\right],
 \quad
 t_b = \frac{(1-b)c}{1-bc}t;
 \label{eqn66}
\end{equation}
the formulas for $s_c$ and $t_c$ are obtained by interchanging $b$ and
$c$. A density operator of the form $\rho_R$ has eigenvalues $1-2s, s+t, s-t$,
and those of its image under $\BC$ are the same with $s$ and $t$ replaced by
$s_b$ and $t_b$, and similarly for $\CC$. This leads to comparatively simple
formulas for the von Neumann entropy and thus $\Dl S(\rho)$, so much of
the analysis can be carried out with explicit formulas.

As $b$ increases out of the degradable region to the left of line 1 in
Fig.~\ref{fgr2} the breakdown of the concavity of $\Dl S(\rho)$ is connected
with a change in sign of $[\pr^2 \Dl S(\rho_R) /\pr t^2]_{t=0}$. Equating this
quantity to zero yields a formula
\begin{equation}
s = \frac{(1-bc)(b+c-2bc)}{(-3bc)(1+2b+2c-2bc)}
\label{eqn67}
\end{equation}
for $s$ as a function of $b$ and $c$. Setting $s=0$ gives the formula
\begin{equation}
 c = -b/(1-2b)
\label{eqn68}
\end{equation}
for the dotted line, and $\Dl S(\rho)$ is concave everywhere to
the left of this line. As $b$ increases further the value of $s$ given by
\eqref{eqn67} increases until it reaches the value at which $\Dl S(\rho_0)$
is maximum. Setting $d \Dl S(\rho_0)/d s =0$ yields a formula
\begin{equation}
 (1-b)(1+2b)\, c\, \log(-2+1/s_b) = (1-c)(1+2c)\, b\, \log(-2+1/s_c)
\label{eqn69}
\end{equation}
in which $s_b$ and $s_c$ depend on $s$ through \eqref{eqn66} and its
counterpart with $b$ and $c$ interchanged.  Eliminating $s$ between
\eqref{eqn67} and \eqref{eqn69} gives a formula relating $b$ and $c$ which can
be solved numerically to give the dot-dash curve in the figure.
Line 3 in the figure is where the maximum of $\Dl S(\rho_2)$ goes to zero,
terminating the region where $\QC^{(1)}(\BC)>0$.  This occurs when the
eigenvalues of $\BC(\rho_2)$ as functions of $s$ become degenerate with those of
$\CC(\rho_2)$ along the line
\begin{equation}
 c = -b/(1+b).
\label{eqn70}
\end{equation}

The remaining curves in Fig.~\ref{fgr2} are associated with the complementary
channel $\CC$, with $\QC^{(1)}(\CC)$ given by the minimum rather than the 
maximum of $\Dl S(\rho)$.  To the right of line 4,  $\CC$ is
degradable ($\BC$ antidegradable) and $\Dl S(\rho)$ is a convex function. 
Our numerical studies indicate that this convexity persists for slightly
smaller value of $b$, and first breaks down at $s=0$ when $d^2 \Dl
S(\rho_0)/d s^2$ changes sign. Equating this to zero at $s=0$ produces the
relatively simple expression
\begin{equation}
 b\,(1-c)(1+2c) = (-c)(1-b)(1+2b)
\label{eqn71}
\end{equation}
and thus a formula
\begin{equation}
 c = \left( 1 + 2b -2b^2 -\sqrt{1+4(b+2b^2-2b^3+b^4)} \right)/4b
\label{eqn72}
\end{equation}
for the curve with narrow dashes lying just to the left of line 4.

As $b$ continues to decrease $\Dl S(\rho)$ is minimized by $\rho =\rho_0$, but
with $s$ increasing until it reaches its maximum value of $1/2$, which first
occurs when $d \Dl S(\rho_0)/d s =0$ at $s=1/2$. The transition line is given by
\eqref{eqn69} with $s_b$ and $s_c$ set equal to their values when $s=1/2$.
Solving it numerically for $c$ as a function of $b$ yields the curve with long
dashes in Fig.~\ref{fgr2}. The optimizing (minimizing) density operator for yet
smaller values of $b$ continues to be $\rho_1$, and $\QC^{(1)}(\CC)$ finally
reaches zero at curve 2, where $S(\BC(\rho_1)) = S(\CC(\rho_1))$, or
\begin{align}
 &(1+2b)(1-c) \log[2(1+2b)(1-c)] +(1-b)(2+c)\log[(1-b)(2+c)] = 
 \notag\\
  &(1+2c)(1-b) \log[2(1+2c)(1-b)] +(1-c)(2+b)\log[(1-c)(2+b)].
\label{eqn73}
\end{align}
A numerical solution gives curve 2 in Fig.~\ref{fgr2}. (The slight hook visible
at the foot of the curve is genuine, and reflects the $(1+2c)\log(1+2c)$ term
in \eqref{eqn73}.)

\end{document}